
\baselineskip=16pt
\null\vskip-30pt
\rightline{SWAT/94/47}
\rightline{hep-th/9411016}
\rightline{November 1994}
\vskip0.5 truecm
\centerline{\bf $O(1/N_f)$ CORRECTIONS TO THE THIRRING MODEL IN
$2<d<4$}
\vskip 0.5 truecm
\centerline{{\bf Simon Hands}}
\centerline{\it Department of Physics}
\centerline{\it University of Wales, Swansea,}
\centerline{\it Singleton Park,}
\centerline{\it Swansea SA2 8PP, U.K.}
\vskip 1.5 truecm
\centerline{\bf Abstract}

{\narrower
\noindent
The Thirring model, that is, a relativistic field theory of fermions
with a contact interaction between vector currents, is studied for
dimensionalities $2<d<4$ using the $1/N_f$ expansion, where $N_f$ is the
number of fermion species. The model is found to have no ultraviolet
divergences at leading order provided a regularization respecting
current conservation is used. Explicit $O(1/N_f)$ corrections are
computed, and the model shown to be renormalizable at this order in the
massless limit;
renormalizability appears to hold to all orders due to a special case
of Weinberg's theorem. This implies there is
a universal amplitude for four particle scattering in the asymptotic
regime.
Comparisons are made with both the Gross-Neveu model and QED.

\noindent
PACS numbers: 11.10.Gh, 11.10.Jj, 11.10.Kk, 11.15.Pg
\smallskip}
\baselineskip=20pt
\parskip= 5pt plus 1pt
\parindent=15pt
\vskip 1 truecm
\noindent{\bf I. Four Fermi Theories in $d\in(2,4)$}

It has been believed for some time that a renormalizable expansion for
four-fermion models exists for dimensions larger than two, which is
naively the upper critical dimension [1 - 5]. Rather than
using the coupling constant $g^2$, which has inverse dimensions of mass
for $d>2$, to organize the expansion, the dimensionless parameter
$1/N_f$, where $N_f$ is the number of fermion species, is used. The
standard example is the Gross-Neveu model:
$${\cal L}=\bar\psi_i\partial{\!\!\! /}\,\psi_i-{g^2\over2N_f}
(\bar\psi_i\psi_i)^2.\eqno(1.1)$$
In this case, which has been widely studied [1 - 8],
spontaneous fermion mass generation occurs for values of
$g^2>g_c^2\sim O(\Lambda^{2-d})$, where $\Lambda$ is some ultraviolet
cutoff in the model. It is preferable to discuss the problem with $d$ a
continuous parameter, $d\in(2,4)$; the scaling properties of the model
are then more transparent. Of course, only $d=3$ can correspond to a
physically realizable system. If the coupling is now fine tuned to the
neighbourhood of  $g_c$, then light fermions propagate and interact via
exchange of a composite scalar state of mass $2m$, where
$m$ is the fermion mass. (Actually a
perfectly acceptable model also arises by approaching $g_c$ from the
massless phase). Because the model is strongly interacting at $g=g_c$,
the ultraviolet asymptotic behaviour of the scalar propagator, obtained
by resumming a sequence of fermion -- anti-fermion bubble diagrams which
are dominant at leading order in $1/N_f$, is non-standard:
$$\displaystyle\lim_{k^2\to\infty}D_\sigma(k)\propto{1\over k^{d-2}}.
\eqno(1.2)$$

The behaviour (1.2) of $D_\sigma$ when input to a standard power
counting analysis [4,5] implies that the superficial degree of
divergence of Feynman diagrams describing corrections of higher order in
$1/N_f$ does not depend on the number of interaction vertices, which in
turn suggests that the expansion is exactly renormalizable. This
property has been explicitly verified at $O(1/N_f)$ [5,7 - 9].
Physically, the renormalizability of the model may be understood as a
consequence of its being the infra-red fixed point under renormalization
group flow of a model of fermions interacting with elementary (ie. not
auxiliary) scalar fields via a Yukawa interaction [3,7,10].
This model is super-renormalizable. The
IR fixed point of the Yukawa model is identical to a UV fixed point of
the Gross-Neveu model as $g^2\to g_c^2$. The relation between
renormalizability and hyperscaling relations between the model's
critical exponents (which are polynomials in $1/N_f$) was stressed in
[7,8]. The exponents are currently all known to $O(1/N_f^2)$ [11],
some to $O(1/N_f^3)$ [12], and
have been verified for $d=3$
by numerical simulation first for $N_f=12$ [8] and
most recently for $N_f=2$ [13]. The situation is analogous to that in
interacting scalar field theories; there the IR fixed point of
the super-renormalizable
$\phi^4$ theory in $d\in(2,4)$ is identical to the UV fixed point of a
corresponding nonlinear sigma model [1,3], which once again has an
unexpected renormalizability in $1/N_f$ [14,15].

In this paper I wish to concentrate on another, distinct interacting
fermion theory, a generalization of the massive Thirring model. The
Lagrangian is
$${\cal L}=\bar\psi_i(\partial{\!\!\! /}\,+m)\psi_i+
{g^2\over 2N_f}(\bar\psi_i\gamma_\mu\psi_i)^2.\eqno(1.3)$$
This model has also been studied in the $1/N_f$ expansion [16 - 19].
In this case there is no phase transition corresponding to spontaneous
mass generation, but instead the ``vacuum polarization'' fermion bubble
diagrams correcting the intermediate boson propagator prove to be UV
finite, despite a superficial $\Lambda^{d-2}$ divergence, due to
fermion current conservation. The situation is analogous to QED, where
due to current conservation, the transverse projection operator
$(\delta_{\mu\nu}k^2-k_\mu k_\nu)$ can always be factored from the
vacuum polarization, reducing the effective degree of divergence by
two. In QED the result is that the photon remains massless at each
order of perturbation theory.

It is interesting to contrast the Thirrng model with the Gross-Neveu
model. In the latter
case the superficial $\Lambda^{d-2}$ divergences do not cancel, and
result in an additive renormalization of the coupling $g^2$, and hence
the need to fine-tune to recover the continuum limit.
In the Thirring case, the $\Lambda^{d-2}$ divergences vanish for essentially
kinematic reasons, the coupling $g^2$ is not renormalized (as shown
below), and the continuum limit appears to exist, to leading order in
$1/N_f$, for all values of $g^2$.
As in the Gross-Neveu model, the
expression for the intermediate boson propagator in the Thirring model,
which is now a vector, can be resummed, to give the same asymptotic form
as eqn. (1.2). Therefore the power counting arguments also suggest that
the $1/N_f$ expansion is renormalizable. However, the underlying physics
is different; there is no phase transition corresponding to a fixed
point condition, and apparently no underlying super-renormalizable model which
yields equivalent physics in its IR limit. A proof of renormalizability,
therefore, must lie entirely within the context of the $1/N_f$ expansion
of the four-fermi theory.

The potential problem which might arise for renormalization at higher
orders in four-fermi models was highlighted in [8]. At
next-to-leading order, the corrections to the boson propagator are given
by two-loop diagrams, exemplified in fig. 4. These are superficially
$\Lambda^{d-2}$ divergent, as discussed above, but there are also
subleading divergences of the form (on dimensional grounds)
$k^{d-2}\ln\Lambda$,
divergent contributions which are non-polynomial in external momentum,
and which hence cannot be compensated by the addition of a local
counterterm. These terms, if not removed by explicit cancellation with
other divergent graphs at the same order (which was demonstrated for
the Gross-Neveu model in [8]), would spoil the
renormalizability  of the model, and result in a non-local interaction
being generated as the cutoff is removed. As we shall see, this
constraint on subdivergences is a feature of a graphical expansion in
which diagrams of one and two loops appear at the same order. It will
turn out that the cancellation of non-polynomial divergences is a
natural consequence of Weinberg's theorem [20] applied to theories with
non-standard propagators. There appears to be no barrier to extending a
renormalizability proof to all orders using standard arguments.

The rest of the paper is organized as follows. In section II we review
the work of [18,19] in setting up the $1/N_f$ expansion for the
Thirring model at leading order, extending their work, which was for the
special (but physical!) case $d=3$, to the interval $d\in(2,4)$.
We shall give a closed form expression for the auxiliary vector
propagator, and examine it in various limits, including the important
deep Euclidean limit $k^2\to\infty$. The mass of the resulting vector
boson is discussed as a function of $g$ and $m$. In section III the
divergence structure of the model is discussed, $O(1/N_f)$ corrections
computed, and the renormalization of the model at this order given.
The condition that the fermion current is conserved translates into a
requirement that the vacuum polarization is two-loop finite: this is
verified explicitly. In section IV we compare the Thirring model with
the Gross-Neveu model and show why the cancellation of non-polynomial
divergences is to be expected as a result of Weinberg's theorem:
the result is
that both models have a very similar asymptotic structure corresponding
to an interacting  UV fixed point,
despite the contrast at low energies. Finally,
comparisons are drawn between the Thirring model and QED, and possible
implications for a non-trivial fixed point for the latter are
discussed.
\vskip 1 truecm
\noindent{\bf II. Leading Order Results}

Consider the Lagrangian for the Thirring model in the following
bosonised form:
$${\cal L}=\bar\psi_i\partial{\!\!\! /}\,\psi_i+m\bar\psi_i\psi_i
+{{ig}\over\sqrt{N_f}}A_\mu\bar\psi_i\gamma_\mu\psi_i+{1\over2}A_\mu
A_\mu,\eqno(2.1)$$
where sums on repeated spacetime indices $\mu$ and flavor indices $i$
are understood. The field $A_\mu$ is a vector auxiliary: it may be
integrated over to recover the original Lagrangian (1.3). In
$d$-dimensional Euclidean space, $d\in(2,4)$, we define
$\{\gamma_\mu,\gamma_\nu\}=2\delta_{\mu\nu}1\kern-4pt 1$,
$\delta_{\mu\mu}=d$, ${\rm tr}1\kern-4pt 1=4$; that is, we assume four
component spinors and hence avoid complications due to parity violation
and generation of a Chern-Simons term in $d=3$ [18,19]. The
Feynman rules are thus
$$\eqalign{
{\rm fermion\; propagator}:\;\;\; S_F&=(ip{\!\!\! /}\,+m)^{-1}\cr
{\rm interaction\; vertex}:\;\;\;
\Gamma_\mu&=-{{ig}\over\sqrt{N_f}}\gamma_\mu\cr}\eqno(2.2)$$
To leading order in $1/N_f$ the auxiliary propagator receives a
contribution from ``vacuum polarization'', that is, a fermion --
anti-fermion bubble (figure 1), so we write
$${\rm vector\; propagator}:\;\;\;D_{\mu\nu}^{-1}(k)=\delta_{\mu\nu}-
\Pi_{\mu\nu}(k),\eqno(2.3)$$
where the leading order vacuum polarization is given by
$$\Pi_{\mu\nu}(k)={g^2\over N_f}\times N_f\times
\int_p{\rm tr}\;\gamma_\mu{1\over{i(p{\!\!\! /}\,+ak{\!\!\! /}\,)+m}}
\gamma_\nu{1\over{i(p{\!\!\! /}\,+(a-1)k{\!\!\! /}\,)+m}}.\eqno(2.4)$$
The constant $a$ defining the momentum routing is kept arbitrary at
present. For $d\in(2,4)$ momentum integration is defined as follows:
$$
\int_p f(p^2)=S_d\int_0^\infty p^{d-1}f(p^2)dp,\eqno(2.5)$$
with
$$\int_p p_\mu p_\nu
f(p^2)={\delta_{\mu\nu}\over d}\int_p p^2f(p^2)\;;\;\;\;
\int_p p_\mu p_\nu p_\lambda p_\kappa f(p^2)=
{(\delta_{\mu\nu}\delta_{\lambda\kappa}+\delta_{\mu\lambda}\delta_{\nu\kappa}
+\delta_{\mu\kappa}\delta_{\nu\lambda})\over{d(d+2)}}\int_p p^4 f(p^2)\;;
\;\;\;{\rm etc,}\eqno(2.6)$$
and
$$S_d\equiv{2\over{(4\pi)^{d\over2}\Gamma({d\over2})}}.\eqno(2.7)$$
At this stage no regularization is specified. We now perform the trace
and then apply Schwinger parameterization:
$$\eqalign{
\Pi_{\mu\nu}(k)&=4g^2\int_0^\infty d\alpha d\beta\int_p
\exp\left[-\alpha((p+ak)^2+m^2)-\beta((p+(a-1)k)^2+m^2)\right]\cr
&\times\left[-2p_\mu p_\nu+2a(1-a)k_\mu k_\nu+(1-2a)(p_\mu k_\nu+p_\nu
k_\mu)
+\delta_{\mu\nu}(p^2+a(a-1)k^2+k.p+m^2)\right].\cr}
\eqno(2.8)$$
Since $\int_p$ is now finite, the momentum $p$ may be shifted and the
integral performed; the result is
$$\eqalign{
\Pi_{\mu\nu}(k)&={{4g^2}\over(4\pi)^{d\over2}}\int_0^\infty
{{d\alpha d\beta}\over(\alpha+\beta)^{d\over2}}
\exp\left[-(\alpha+\beta)m^2-{{\alpha\beta}\over(\alpha+\beta)}k^2\right]\cr
&\times\left\{-{{2\alpha\beta}\over(\alpha+\beta)^2}(k^2\delta_{\mu\nu}
-k_\mu k_\nu)
+\delta_{\mu\nu}\left[m^2+{{\alpha\beta}\over(\alpha+\beta)^2}k^2
+{{d-2}\over{2(\alpha+\beta)}}\right]\right\}.\cr}\eqno(2.9)$$
Note that all dependence on the momentum routing parameter $a$ has
disappeared. Now the integral of the second term in curly brackets,
proportional to
$\delta_{\mu\nu}$, may be reexpressed as
$${{\partial\phantom{i}}\over{\partial x}}\int_0^\infty
{{d\alpha d\beta}\over(\alpha+\beta)^{{d\over2}+1}}
{1\over x^{{d\over2}-1}}\exp\left[-x{{\alpha\beta}\over(\alpha+\beta)}k^2
-x(\alpha+\beta)m^2\right]\Biggr\vert_{x=1}.\eqno(2.10)$$
However, it can be seen that the integral in (2.10) is formally
independent of $x$, by rescaling $\alpha$ and $\beta$. Strictly, the
integral diverges and must be made finite
by use of a Pauli-Villars regulator field (eg.
see [21] ch. 7). Its contribution to $\Pi_{\mu\nu}$ thus vanishes, and we can
write
$$\Pi_{\mu\nu}={\cal P}_{\mu\nu}(k)\Pi(k^2),\eqno(2.11)$$
with the transverse projection operator ${\cal P}_{\mu\nu}(k)$ defined
as
$${\cal P}_{\mu\nu}(k)=\left(\delta_{\mu\nu}-{{k_\mu k_\nu}\over
k^2}\right).\eqno(2.12)$$
The remaining integrals over $\alpha$ and $\beta$ are straightforward,
and the result for $\Pi(k^2)$ is finite:
$$\eqalign{
\Pi(k^2)&=-{{8g^2k^2}\over(4\pi)^{d\over2}}\Gamma(2-{\textstyle{{d\over2}}})
\int_0^1dx{{x(1-x)}\over[x(1-x)k^2+m^2]^{2-{d\over2}}}\cr
&=-g^2{{4\Gamma(2-{d\over2})}\over{3(4\pi)^{d\over2}m^{4-d}}}
k^2F(2;2-{\textstyle{d\over2}};{\textstyle{5\over2}};-{k^2\over 4m^2}),\cr}
\eqno(2.13)$$
where $F$ is the hypergeometric function.

For $d<4$, we see that the vacuum polarization tensor $\Pi_{\mu\nu}(k)$ can be
evaluated exactly at leading order, with the assumption of a
regularization which respects current conservation. It is interesting to
compare (2.13) with known results in $d=4$ and $d=2$. In the limit
$d\to4_-$,
$$\Pi_{\mu\nu}(k)=-{g^2\over{6\pi^2}}
\left({1\over(4-d)}+{\gamma_E\over2}\right)
k^2{\cal
P}_{\mu\nu}(k),\eqno(2.14)$$
where $\gamma_E$ is the Euler constant. This is almost the textbook
result for one-loop vacuum polarization in dimensionally regularized
${\rm QED}_4$, except that since we have had no need to introduce a
renormalization scale to make the coupling $g^2$ dimensionless, then there is
no term in $\ln k^2$. If we use the linear transformation
properties of $F$ to examine the limit $m^2\to0$ (eg. [22] ch. 15),
it is also possible subsequently to take the limit $d\to2_+$ with the result
$$\Pi_{\mu\nu}(k)=-2{g^2\over\pi}{\cal P}_{\mu\nu}(k).\eqno(2.15)$$
This is exactly twice the result for the one-flavor massless Schwinger
model (in that case figure 1 generates a dynamical photon mass
$g/\sqrt{\pi}$); the extra factor of 2 arises from our insistence on
four component spinors.

Now we set $d=3$ to obtain
$$\Pi_{\mu\nu}(k)=-{g^2\over{2\pi}}{\cal P}_{\mu\nu}(k)
\left[m+{1\over{2(k^2)^{1\over2}}}(k^2-4m^2)\tan^{-1}
\left({{(k^2)^{1\over2}}\over{2m}}\right)\right].\eqno(2.15)$$
This is identical to the result of Yang [18] and Gomes {\it et al\/} [19],
with the momentum $k$ analytically continued to Euclidean space. It is
important to note that for the whole range $d\in(2,4)$ the asymptotic
form of $\Pi_{\mu\nu}(k)$ is not polynomial in $k^2$, viz:
$$\displaystyle\lim_{k^2\to\infty}\Pi_{\mu\nu}(k)=
-g^2{\cal P}_{\mu\nu}(k){(k^2)^{{d\over2}-1}\over A_d},\eqno(2.16)$$
with the numerical constant $A_d$ given by
$$A_d={{(d-1)}\over{(d-2)}}
{(4\pi)^{d\over2}\over{4\Gamma(2-{d\over2})B({d\over2},{d\over2}-1)}},
\eqno(2.17)$$
where $B$ is the Beta function. The form (2.16) was first found by
Hikami and Muta [17], modulo a difference of definition of
${\rm tr}1\kern-4pt 1$, and a factor of $d$.

Let us now return to expression (2.3) for the inverse vector propagator. Using
(2.11) we can invert to yield the propagator
$$D_{\mu\nu}(k)={\cal P}_{\mu\nu}(k){1\over{1-\Pi(k^2)}}+{{k_\mu k_\nu}\over
k^2}.\eqno(2.18)$$
As argued in [16,18,19], the second term in $D_{\mu\nu}(k)$, which is
longitudinal, behaves as a constant in the limit $k\to\infty$, and might
naively be expected to lead to poor ultraviolet behaviour. However, since
the vector auxiliary interacts with a physical current which is
conserved, S-matrix elements and observables constructed as
gauge-invariant combinations of the fields $\psi$ and $A_\mu$
(in the sense used in QED) should not
be affected by this problem, although Green functions in general might
be. We shall see this when we calculate $O(1/N_f)$ corrections in the
next section. To aid calculation (but not to
{\sl define\/} the vector propagator, as would be the case in QED),
following [19], we introduce a
gauge-fixing term ${1\over{2\mu^2}}(\partial_\mu A_\mu)^2$ to the
Lagrangian (2.1), which has the effect of moderating the UV behaviour of
the second term for finite $\mu$, which has dimensions of mass. The
vector propagator becomes
$$D_{\mu\nu}(k;\mu)={\cal P}_{\mu\nu}(k){1\over{1-\Pi(k^2)}}
+{\mu^2\over{\mu^2+k^2}}{{k_\mu k_\nu}\over
k^2}.\eqno(2.19)$$
The scale $\mu$ is in effect a regulator which should not appear in
final expressions. The limit $\mu\to\infty$ recovers the original
Thirring model, whereas the limit $\mu\to0$ specifies a ``Landau gauge''
which will render the two-loop calculation in the next section much
easier. Note also that in the limit $k^2\to0$,
$\Pi(k^2)\to0$ and $D_{\mu\nu}(k)\to\delta_{\mu\nu}$: hence the
infrared problems associated with QED are not present here.

On the assumption that the longitudinal piece of $D_{\mu\nu}$ has no
physical consequence, we focus on the transverse piece and identify
a pole condition for the mass of the vector $M_V$:
$$1-g^2M_V^2
{{4\Gamma(2-{d\over2})}\over{3(4\pi)^{d\over2}m^{4-d}}}
F(2;2-{\textstyle{d\over2}};{\textstyle{5\over2}};{M_V^2\over 4m^2})=0.
\eqno(2.20)$$
In general this is a transcendental equation. It can be solved in two
limits. For strong coupling $g^2\gg m^{2-d}$
the vector channel will be dominated by a tightly bound
fermion -- anti-fermion state, so $M_V^2\ll m^2$. Therefore we can
expand $F$ to obtain
$${M_V^2\over m^2}\simeq{m^{2-d}\over g^2}\left({{3(4\pi)^{d\over2}}\over
{4\Gamma(2-{d\over2})}}\right).\eqno(2.21)$$
For arbitrarily weak coupling, real solutions of (2.20) can only exist
if the hypergeometric function is able to grow arbitrarily large. We expect a
weakly bound state to have mass given by
$$M_V=2m-\varepsilon.\eqno(2.22)$$
As $M_V\to(2m)_-$, the hypergeometric function diverges only for $d<3$.
In this case we can once again perform a linear transformation on $F$ to
get
$$1-g^2M_V^2{{\sqrt{\pi}\Gamma({{3-d}\over2})}\over{m^{4-d}(4\pi)^{d\over2}}}
\left({\varepsilon\over
m}\right)^{{d-3}\over2}\left(1+O\left({\varepsilon\over m}\right)\right)=0,
\eqno(2.23)$$
ie. the binding energy is
$$\varepsilon=m\left(g^2m^{d-2}{{\Gamma({{3-d}\over2})}\over
{2^{d-2}\pi^{{d-1}\over2}}}\right)^{\textstyle{2\over{3-d}}}.\eqno(2.24)$$
The case $d=3$ must be handled separately; we use expression (2.15) to
find the binding energy is essentially singular in $g^2$:
$$\varepsilon=2m\exp\left(-{{2\pi}\over{mg^2}}\right).\eqno(2.25)$$
Finally, for $d>3$ the hypergeometric function remains finite as
$k^2\to-4m^2$. In this case the bound state vanishes (ie. the would-be pole
coalesces with the branch cut in $F$) for values of $g$
below a critical $g_c$ given by
$$g_c^2=m^{2-d}{{(4\pi)^{d\over2}(d-1)(d-3)}\over{16\Gamma(2-{d\over2})}}.
\eqno(2.26)$$
In the subcritical region $D_{\mu\nu}$ has no poles on the physical
sheet; the vector can only be regarded as a resonant
intermediate state in four fermion scattering.

In the deep Euclidean region $k^2\to\infty$ things simplify
considerably: the vector propagator has the form
$$\displaystyle\lim_{k^2\to\infty} D_{\mu\nu}(k)=
{\cal P}_{\mu\nu}(k){A_d\over
g^2}{1\over(k^2)^{{d\over2}-1}}.\eqno(2.27)$$
In this limit the four fermion scattering amplitude has the form
$A_d J_\mu(q)J_\mu(q+k)/N_fk^{d-2}$. As we shall argue in the final
section, this interaction receives no corrections in the $1/N_f$
expansion, and is thus a universal form characterizing the short
distance structure
of the model; in other words it defines a UV fixed point. In this respect it
resembles the Gross-Neveu model as discussed in [8] (though note the
definition of $A_d$ is distinct). In the next section when the
renormalization of the model at $O(1/N_f)$ is discussed, the form (2.27)
will be used throughout.
\vskip 1 truecm
\noindent{\bf III. Renormalization at $O(1/N_f)$}

In this section I will discuss the renormalization of the model to
next-to-leading order in the $1/N_f$ expansion. First let me review why
we might expect such a programme to be feasible. The short distance
fluctuations of the model are encoded in the asymptotic form for the
vector propagator (2.27). Suppose we analyze the superficial degree of
divergence of a higher order diagram with $N_\psi$ external fermion
lines, $N_A$ external auxiliary lines, and $V$ vertices. If we use the
form (2.27), then standard power counting analysis gives the
superficial degree of divergence $\omega$:
$$\omega=d-{{d-1}\over2}N_\psi-N_A.\eqno(3.1)$$
It is interesting to compare this result with that for canonical boson
asymptotics, $D(k)\sim1/k^2$, which applies, say, for QED:
$$\omega_{can}=d-{{d-1}\over2}N_\psi-{{d-2}\over2}N_A-{{4-d}\over2}V.
\eqno(3.2)$$
For $d<4$ the degree of divergence falls as the number of vertices
increases: this is characteristic of a super-renormalizable theory. Only
when $\omega$ is independent of $V$ can a perturbative expansion be
exactly renormalizable (ie. divergent graphs appear at every order of
the expansion, but can always be made finite by retuning a
finite set of counterterms).

Using (3.1) we can compile a list of potentially dangerous graphs for
$d\in(2,4)$. The $O(1/N_f)$ contributions are shown in figures 2 - 4.
The fermion self-energy $\psi\psi$
(fig. 2) has $\omega=1$, as in ${\rm QED}_4$,
but since that divergence is odd in loop momentum, the true divergence
is logarithmic ($\omega=0$). The vertex correction
$\psi\psi A$ (fig. 3) also has
$\omega=0$, but the four-vector $AAAA$ scattering, which is superficially
divergent in ${\rm QED}_4$, here has $\omega=d-4$ and so is safe.
One-point and three-point vector scattering vanish by Furry's theorem,
leaving the vector two-point function $AA$ (fig. 4), with $\omega=d-2$ as
the only other superficially divergent case. One can then argue [19]
that in a regularization which respects current conservation, one can
always extract a factor $k^2{\cal P}_{\mu\nu}(k)$ from these diagrams
to give $\omega=d-4$, and hence no new divergence. The model (2.1) can
then be renormalized simply by rescaling the $\psi$ and $A$ fields,
and retuning the fermion mass. I
shall show in this section that this conclusion is correct,
though the argument is not quite so straightforward. Because
the vacuum polarization is non-polynomial in $k^2$, it is in fact only
permissible to extract ${\cal P}_{\mu\nu}(k)$ from inside the graph,
which does not improve the power-counting. As we shall see, there are
divergent contributions both of degree $\omega=d-2$ and $\omega=0$.
The applicability of power counting to four fermion models is
discussed further in the final section.

It is worth contrasting the four-fermi case with the situation in pure
scalar theories. In the renormalization of the nonlinear sigma model in
$d\in(2,4)$ [15], the power counting gives
$$\omega=d-{{d-2}\over2}N_\psi-2N_A,\eqno(3.3)$$
where now $\psi$ denotes the elementary scalar field and $A$ the
auxiliary scalar boson. The only superficial divergences are
$\psi\psi$ ($\omega=2$) and $\psi\psi A$ ($\omega=0$); the auxiliary
propagator $AA$ has $\omega=d-4$ and hence is superficially convergent.
On dimensional grounds there is no reason to expect any non-polynomial
divergences.

The procedure for renormalizing the model follows the treatment in
[8]: first we redefine the Lagrangian
$${\cal L}=Z_\psi\bar\psi_i(\partial{\!\!\! /}\,+M)\psi_i
+{{ig}\over\sqrt{N_f}}Z_\psi Z_A^{1\over2}
\bar\psi_i A{\!\!\! /}\,\psi_i+{1\over2}Z_A\left(A_\mu^2+{1\over\mu^2}
(\partial_\mu A_\mu)^2\right).
\eqno(3.4)$$
The constants $Z_\psi$, $Z_A$, $M$ and in principle $g$ and $\mu$ are
all cutoff-dependent, and must be adjusted at each order of the $1/N_f$
expansion to keep physical matrix elements finite. As we have seen, at
leading order an adequate choice is: $Z_\psi=1$; $M=m$, the physical
fermion mass; $\mu\to\infty$; and $g$, $Z_A$ unconstrained. $Z_A$
simply defines the scale of an auxiliary field at leading order and
hence has no physical relevance. The first divergent Green function we
must examine is the fermion self-energy (fig. 2):
$$\Sigma(k)=-{g^2\over N_f}{{Z_\psi^2Z_A}\over{Z_\psi Z_A}}\int_p
\gamma_\mu{1\over{i(p{\!\!\! /}\,+k{\!\!\! /}\,)+M}}\gamma_\nu
D_{\mu\nu}(p;\mu).\eqno(3.5)$$
Note that with the definition (3.4), the vector propagator at leading
order is $Z_A^{-1}D_{\mu\nu}(k;\mu)$, with $D_{\mu\nu}$ defined by
(2.19). On rearranging, we find
$$\Sigma(k)=-{g^2\over N_f}Z_\psi\int_p\gamma_\mu
{(-i(p{\!\!\! /}\,+k{\!\!\! /}\,)+M)\over{(p+k)^2+M^2}}\gamma_\nu
\left[A(p^2){\cal P}_{\mu\nu}(p)+B(p^2;\mu){{p_\mu p_\nu}\over
p^2}\right],\eqno(3.6)$$
with
$$\displaystyle\lim_{k^2\to\infty}A(k^2)={A_d\over{g^2(k^2)^{{d\over2}-1}}}
\;\;\;;\;\;\;B(k^2)={\mu^2\over{\mu^2+k^2}}.\eqno(3.7)$$

We will treat the parts depending on $A(p^2)$ and $B(p^2;\mu)$
separately. For the first piece, apart from a term which is odd in $p$
and hence vanishes on $\int_p$, the leading contribution is
$O(p^{-1}dp)$, and hence logarithmically divergent. With the choice of
a simple momentum cutoff $|p|<\Lambda$, and the definitions (2.5,6), we
find
$$\Sigma^A(k)=-Z_\psi{C_d\over N_f}{(d-1)^2\over{2(d-2)}}
\left[ik{\!\!\! /}\,{(d-4)\over d}+M\right]\ln{\Lambda\over
M}+{\rm finite},\eqno(3.8)$$
with the constant $C_d$ defined, as in [8]:
$$C_d={1\over{B({d\over2},2-{d\over2})B({d\over2},{d\over2}-1)}}.\eqno(3.9)$$
Note $C_d$ is positive definite for $d\in(2,4)$. For $d=3$, $C_d$
has the value $4/\pi^2$. The piece depending on
$B(p^2;\mu)$ can be recast in the form
$$\Sigma^B(k)=-{{g^2Z_\psi\mu^2}\over
N_f}{\Gamma(2-{d\over2})\over(4\pi)^{d\over2}}\int_0^1dx
[x(1-x)k^2+xM^2+(1-x)\mu^2]^{{d\over2}-2}
\left[i(1+x)k{\!\!\! /}\,+M-{{2i}\over d}k{\!\!\!
/}\,\right]+O(\mu^{d-6}).\eqno(3.10)$$
In the limit $\mu^2\to\infty$:
$$\Sigma^B(k)=-{{g^2Z_\psi\mu^{d-2}}\over N_f}
{\Gamma(2-{d\over2})\over(4\pi)^{d\over2}}{2\over(d-2)}
(ik{\!\!\! /}\,+M)+O(\mu^{d-4}).\eqno(3.11)$$
Note that $\Sigma^A$ and $\Sigma^B$ have different characteristics:
$\Sigma^A$ is cutoff-dependent but independent of the coupling constant
$g$, whereas $\Sigma^B$ is finite for finite $\mu$, but depends on both
$\mu$ and $g$. The Thirring model limit $\mu^2\to\infty$ cannot be
taken for $\Sigma$ in isolation.

We can now write for the full inverse fermion propagator
$$S_F^{-1}(k)
=Z_\psi(ik{\!\!\!
/}\,+M-\Sigma(k))
\equiv ik{\!\!\! /}\,+m,\eqno(3.12)$$
thus defining the wavefunction renormalization constant $Z_\psi$ and
the renormalized (ie. physical) mass $m$ in terms of the bare mass
$M$:
$$\eqalignno{
Z_\psi&=1-{C_d\over N_f}{{(d-1)^2(d-4)}\over{2d(d-2)}}\ln{\Lambda\over
m}-{{g^2\mu^{d-2}}\over N_f}{{2\Gamma(2-{d\over2})}\over
{(4\pi)^{d\over2}(d-2)}}+{\rm finite},&(3.13)\cr
m&=M\left(1+{C_d\over N_f}{{2(d-1)^2}\over{d(d-2)}}\ln{\Lambda\over m}
+{\rm finite}\right)>M.&(3.14)\cr}$$
Thus we find an expression for the physical mass $m$ independent of the
regulator $\mu$, and a wavefunction constant $Z_\psi$ which depends on
$\mu$. This is, of course, very similar to what is found for QED: the
physical parameter $m$ is renormalized in a gauge-invariant fashion,
whereas the unphysical $Z_\psi$ is not. The term proportional to $C_d$
in (3.13) was derived in [17], but the $\mu$-dependent piece was
neglected.

Next we calculate the $O(1/N_f)$ contribution $\Gamma^{[1]}_\lambda$ to the
vertex (fig. 3). For zero external momentum we have
$$\eqalign{
\Gamma^{[1]}_\lambda&={{ig^3}\over N_f^{3\over2}}
{{Z_\psi^3Z_A^{3\over2}}\over{Z_\psi^2Z_A}}\int_p\gamma_\mu
{1\over{ip{\!\!\! /}\,+M}}\gamma_\lambda
{1\over{ip{\!\!\! /}\,+M}}\gamma_\nu D_{\mu\nu}(p;\mu)\cr
&\simeq-ig{{Z_\psi Z_A^{1\over2}}\over\sqrt{N_f}}{g^2\over N_f}\int_p
{1\over(p^2+M^2)^2}D_{\mu\nu}(p;\mu)\gamma_\mu p{\!\!\! /}\,
\gamma_\lambda p{\!\!\! /}\,\gamma_\nu,\cr}\eqno(3.15)$$
where the second line follows because we are only interested in the
divergent part. Using the same procedure as before, we find for the
full vertex
$$\eqalign{
\Gamma_\lambda&=\Gamma_\lambda^{[0]}+\Gamma_\lambda^{[1]}\cr
&=-ig{{Z_\psi Z_A^{1\over2}}\over\sqrt{N_f}}\gamma_\lambda
\left[1+{1\over
N_f}\left(C_d{{(d-1)^2(d-4)}\over{2d(d-2)}}\ln{\Lambda\over m}+
g^2\mu^{d-2}{{2\Gamma(2-{d\over2})}\over{(4\pi)^{d\over2}(d-2)}}\right)
+{\rm finite}\right].\cr}\eqno(3.16)$$
However, the constant $Z_\psi$ has already been determined in (3.13),
and is found to exactly cancel both $\Lambda$- and $\mu$-dependent
terms in (3.16). Hence
$$\Gamma_\lambda=-{ig\over\sqrt{N_f}}Z_A^{1\over2}\gamma_\lambda
\left[1+O(1/N_f)\times{\rm finite}\right].\eqno(3.17)$$
Once again, this is a familiar situation from QED: current conservation
plus gauge invariance ensures that the divergent and gauge-dependent
parts of the self-energy and vertex corrections cancel, ie. $Z_1=Z_2$
(as noted in section II, there are no problems with infrared
divergences in the Thirring case). We expect this cancellation to
persist at higher orders. So, to maintain the finiteness of fermion
self-energy and vertex corrections to $O(1/N_f)$, our only requirement
of $Z_A$ and $g$ so far is that the combination $Z_A^{1\over2}g$ be
finite. However, we have not yet exhausted the list of superficially
divergent graphs. We next consider the $O(1/N_f)$ corrections to the
vector propagator, which consists of two two-loop diagrams (figs. 4a,b):
$$\eqalign{
\Pi^{[1]}_{\mu\nu}(k)&=-{g^4\over N_f^2}\times{{Z_\psi^4Z_A^2}\over
{Z_\psi^4Z_A}}\times
N_f\times\left[2I_{\mu\nu}^a(k)+I_{\mu\nu}^b(k)\right];\cr
I_{\mu\nu}^a(k)&=\int_{pq}{\rm tr}\left(\gamma_\mu{1\over{ip{\!\!\!
/}\,}}\gamma_\alpha{1\over{i(p{\!\!\! /}\,+q{\!\!\! /}\,)}}\gamma_\beta
{1\over{ip{\!\!\! /}\,}}\gamma_\nu{1\over{i(p{\!\!\! /}\,+k{\!\!\!
/}\,)}}\right)D_{\alpha\beta}(q),\cr
I_{\mu\nu}^b(k)&=\int_{pq}{\rm tr}\left(\gamma_\mu{1\over{ip{\!\!\!
/}\,}}\gamma_\alpha{1\over{i(p{\!\!\! /}\,+q{\!\!\! /}\,)}}\gamma_\nu
{1\over{i(p{\!\!\! /}\,+q{\!\!\! /}\,+k{\!\!\! /}\,)}}
\gamma_\beta{1\over{i(p{\!\!\! /}\,+k{\!\!\!
/}\,)}}\right)D_{\alpha\beta}(q),\cr}\eqno(3.18)$$
The fermion masses, which have little impact on the ultraviolet behaviour
of the integrands, have been neglected (see below). Now, given the asymptotic
form (2.27) for $D_{\alpha\beta}(q)\sim A_d/g^2q^{d-2}$, it is easy to
see that the combination of renormalization constants multiplying
$I_{\mu\nu}^{a,b}$ reduces to $Z_Ag^2$, which from previous
considerations must be cutoff-independent. The conclusion is that
$\Pi_{\mu\nu}^{[1]}$ must be UV finite if the model is to be
consistently renormalized at this order. The constant $Z_A$, being
just the scale of an auxiliary field, is undetermined in the model,
and the value of $g$ also appears to be irrelevant as regards the
UV behaviour of the model (Yang [18] points out that $g$ cannot be
renormalized, since it appears in the Lagrangian (3.4) in a
gauge-variant manner after rescaling $A_\mu\mapsto A_\mu/g$).

Now, as shown in [8], diagrams such as those of fig. 4 generically
have divergences of two forms, one independent of $k$, proportional to
$\Lambda^{d-2}+{\rm const}\times M^{d-2}\ln\Lambda$, and the other proportional
to $k^{d-2}\ln\Lambda$. We must show that for the Thirring model both
occur with zero coefficient. First we deal with the momentum-independent
piece, following the technique used in the appendix of [5].

Consider the expression (2.4) for the one-loop vacuum polarization
$\Pi_{\mu\nu}^{[0]}(k)$, and in particular the result of differentiating
it with respect to external momentum $k$. By using the identity
$${{\partial\phantom{k}}\over{\partial k_\mu}}
{1\over{i(p{\!\!\! /}\,+k{\!\!\! /}\,)+M}}=
-{1\over{i(p{\!\!\! /}\,+k{\!\!\! /}\,)+M}}\gamma_\mu
{1\over{i(p{\!\!\! /}\,+k{\!\!\! /}\,)+M}},\eqno(3.19)$$
we see that each differentiation is equivalent to a zero momentum
insertion of $-iA_\mu$ (modulo a factor of $g/\sqrt{N_f}$). Thus
$${{\partial^2\phantom{k}}\over{\partial k_\mu\partial k_\nu}}
\Pi_{\alpha\beta}^{[0]}(k)=-(2(1-a)^2+2a^2)J_{\mu\nu\alpha\beta}^a(k)
+2a(1-a)J_{\mu\nu\alpha\beta}^b(k),\eqno(3.20)$$
where $J_{\mu\nu\alpha\beta}^{a,b}(k)$ are represented in fig. 5.
However, as we showed in section II, $\Pi_{\alpha\beta}^{[0]}$, and hence
$J_{\mu\nu\alpha\beta}^{a,b}$, are finite analytic functions of $k$
which are independent of the momentum routing $a$. Hence
$$0\equiv{{\partial\phantom{i}}\over{\partial a}}
{{\partial^2\phantom{k}}\over{\partial k_\mu\partial k_\nu}}
\Pi_{\alpha\beta}^{[0]}(k)=2(1-2a)[2J_{\mu\nu\alpha\beta}^a(k)
+J_{\mu\nu\alpha\beta}^b(k)].\eqno(3.21)$$
Now we contract the R.H.S. of (3.21) with $D_{\alpha\beta}(k)$ and
perform $\int_k$ to obtain the two-loop integrals $I_{\mu\nu}^{a,b}$
at zero momentum. Since the R.H.S. must vanish for arbitrary $a$,
we conclude
$$2I_{\mu\nu}^a(0)+I_{\mu\nu}^b(0)=0.\eqno(3.22)$$
This argument shows that the momentum-independent part of
$\Pi_{\mu\nu}^{[1]}(k)$ is identically zero, and hence that we do not
need to worry about $\Lambda^{d-2}$ divergences. However, each
diagram is individually divergent, as can be seen by setting $a=0$
in (3.20), then evaluating
$\int_k J_{\mu\nu\alpha\beta}^a(k)D_{\alpha\beta}(k)$; we find for the
leading divergence
$$2\Pi_{\mu\nu}^{[1a]}(0)=\delta_{\mu\nu}
{{4\Lambda^{d-2}}\over{(4\pi)^{d\over2}\Gamma({d\over2})}}
{{(d-1)^2(d-3)}\over{d(d-2)}},\eqno(3.23)$$
where $\Lambda$ is a momentum cutoff for $\int_k$.

Of course, we might have anticipated that the $\Lambda^{d-2}$
superficial divergence vanishes due to current conservation, as
it did at leading order, and indeed as it does in QED to all
orders, where $\Pi_{\mu\nu}(0)=0$ ensures that the photon
propagator retains a zero mass pole, as required by gauge invariance.
However, no such argument constrains the momentum-dependent
divergence $k^{d-2}\ln\Lambda$ -- for instance, in ${\rm QED}_4$,
$k^2\ln\Lambda$
divergences are physical and responsible for charge renormalization -- and it
is
not {\sl a priori\/} clear what will happen in a model with
non-standard asymptotics in $d<4$. To examine this case it is necessary
to perform an explicit two-loop calculation, using the techniques
developed in [8], which are  now outlined.

After performing the trace over Dirac indices, each integral may be
reduced to several components of the form
$${\rm constant}\times\int_q\int_p\int_{\Sigma_i x_i}
{{P_{\mu\nu\alpha\beta}(q,k;x_i)D_{\alpha\beta}(q;\mu)}\over
{[p^2+Q(q,k;x_i)]^n}},\eqno(3.24)$$
using Feynman parameterization and momentum shift in $p$.
Here $P$ and $Q$ are polynomial
functions of momenta, $x_i$ are Feynman parameters which are to be
integrated over a domain $\Sigma_i x_i\leq1$, and $n$ is integer.
The algebra is considerably simplified by the choice of ``Landau
gauge'' $\mu^2\to0$, which means that all terms proportional to
$q_\alpha$, $q_\beta$ can be discarded. The integral over the fermion
loop momentum $p$ can always be performed for $n\geq2$:
$$\int_p{1\over[p^2+Q]^n}={{Q^{{d\over2}-n}\Gamma(n-{d\over2})}\over
{\Gamma(n)(4\pi)^{d\over2}}}.\eqno(3.25)$$
Momentum dependent divergences arise when there are two or more
Feynman parameters, in which case there is an intermediate integral
of the form
$$\int_0^{1-\Sigma_{j\not=i}x_j}dx_i{{A+Bx_i+Cx_i^2+\dots}\over
{(a+bx_i+cx_i^2)^s}},\eqno(3.26)$$
where $s$ is non-integer, and the coefficients $b$, $c$ are
$O(q^2)$, where $q$ is the remaining loop momentum,
but the coefficient $a$ is $O(q^0k^2)$. The contribution to (3.26)
from the $x_i\to0$ limit of the integral is then
$$-{{Ab}\over{(s-1)\Delta a^{s-1}}}\left(1+{(3-2s)\over(2-s)}
{{2ac}\over\Delta}\right)-{B\over{(s-1)(s-2)\Delta a^{s-2}}}+
O\left({1\over q^6}\right),\eqno(3.27)$$
with $\Delta\equiv4ac-b^2$. The exact expression on which this
approximation is based is given in the appendix of [8]. Note
that all factors, including $\Delta$, must be
expanded consistently to $O(1/q^4)$ due to the presence of $O(q^4)$
terms in the numerator polynomial $P$. Once this limit has been
isolated, the remaining integral over $q$ is of the form
$$\int_q{{R_{\alpha\beta\mu\nu}(k,q)k^{d-4}}\over q^2}D_{\alpha\beta}(q),
\eqno(3.28)$$
where $R$ is $O(k^2q^0)$. The form (2.27) for $D_{\alpha\beta}$
is now sufficient to evaluate $\int_q$ with a momentum cutoff: it
yields a logarithmic divergence.
Any remaining integrals over Feynman parameters give combinations of
Beta functions in $d$. The final result is
$$\eqalign{
2\Pi_{\mu\nu}^{[1a]}(k)&=-\Pi_{\mu\nu}^{[1b]}(k)\cr
&={{g^2Z_A}\over N_f}{{\Gamma(2-{d\over2})\Gamma({d\over2})A_d}\over
{(4\pi)^d\Gamma(d)}}{{32(d-1)(d-4)}\over d}{\cal P}_{\mu\nu}(k)
(k^2)^{{d\over2}-1}\ln{\Lambda\over k}\cr
&=g^2Z_A{\cal P}_{\mu\nu}(k){{(k^2)^{{d\over2}-1}}\over A_d}
\left({C_d\over N_f}{{(d-1)^2(d-4)}\over{d(d-2)}}\ln{\Lambda\over k}
\right)\cr}\eqno(3.29)$$
on rearranging. We note that each diagram is individually transverse
and equal and opposite respectively to twice the $\mu$ -independent
parts of the
self-energy correction (3.13) ($a$), or the vertex correction (3.16)
($b$). The main result, of course, is that the two-loop vacuum
polarization $\Pi_{\mu\nu}^{[1]}(k)$ is UV finite in the massless
theory. Strictly,
we have not demonstrated the independence of this result on the ``gauge
fixing'' parameter $\mu$, and must rely on reasoning that the vacuum
polarization, which yields charge renormalization in QED, is
gauge invariant. Perhaps more importantly, we have not considered
divergences of the form $M^{d-4}k^2\ln\Lambda$: these probably exist and
correct the mass of the vector $M_V$ following the discussion
(2.20-26).

We have now exhausted the list of divergent Green functions at
$O(1/N_f)$, and proven that the model is renormalizable at this order
(at least in the massless limit) -- indeed the
only physical (ie. gauge invariant) renormalization that must be made
is that of the fermion mass (3.14).
\vskip 1 truecm
\noindent{\bf IV. Discussion}

We begin the final section by thinking about why the cancellation of
$k^{d-2}\ln\Lambda$ divergences in the two-loop vacuum polarization
takes place. The way the calculation has been presented here suggests
a cancellation between the two diagrams $\Pi_{\mu\nu}^{[1a]}$ and
$\Pi_{\mu\nu}^{[1b]}$ (fig. 4). However, a similar analysis of the
Gross-Neveu model (1.1) [8] suggests it is more natural to think of
a cancellation between $\Pi_{\mu\nu}^{[1a]}$ and twice
the self-energy (3.13),
and between $\Pi_{\mu\nu}^{[1b]}$ and twice the vertex correction
(3.16). For
the analogous diagrams in the Gross-Neveu case (note the constant
$A_d$ is different):
$$\eqalign{
\Pi^{[1a]}(k)&=-2{(k^2)^{{d\over2}-1}\over A_d}{{\partial\Sigma^{[1]}(k)}
\over{\partial(ik{\!\!\! /}\,)}}={(k^2)^{{d\over2}-1}\over A_d}
{(d-2)\over d}{C_d\over N_f}\ln{\Lambda\over k};\cr
\Pi^{[1b]}(k)&=-2{(k^2)^{{d\over2}-1}\over A_d}{{\Gamma^{[1]}(k)}\over
{(-g/\sqrt{N_f})}}={(k^2)^{{d\over2}-1}\over A_d}
{C_d\over N_f}\ln{\Lambda\over k}.\cr}\eqno(4.1)$$
In fact, what has occured is a cancellation between the subdivergences of
figure 4 and the diagrams which would result from inserting local
counterterms arising from the divergences of eqns. (3.13) and (3.16) in
the leading order vacuum polarization figure 1 -- factors of two come
because there are two fermion lines and two vertices to correct.

In a treatment of renormalization which proceeds by subtraction of divergent
parts in an ordered fashion -- the BPHZ scheme -- this cancellation is
transparent. In the presentation here I have chosen a physically more
intuitive approach -- rescaling the fields and coupling parameters in
the bare Lagrangian to keep Green functions finite at each order -- but of
course the two schemes are completely equivalent. However, it is important
to note that the argument depends on a novel application of
Weinberg's theorem [20]. The theorem states, in effect, that a 1PI graph
such as $\Pi_{\mu\nu}^{[1]}$ with positive degree of divergence $\omega$
will be balanced by an overall counterterm which is polynomial in external
momentum (in this case $O(\Lambda^{d-2}k^0)$), provided that its
subdivergences are first subtracted, in this case by applying the
appropriate vertex and self-energy corrections. These subdivergences may
be non-polynomial in momentum, but the theorem guarantees that the
cancellation will be exact. The crucial point is that Weinberg proved the
theorem for a wide class of integrands; the usual integrals built from
the standard Feynman propagators proportional to
$(p^2+m^2)^{-1}$, $(ip{\!\!\! /}\,+m)^{-1}$ etc. form just one class.
There is no general requirement that the
propagators be algebraic in the internal momenta, and integrals including
the resummed propagator $\propto k^{2-d}$ are also included.\footnote
{$^\dagger$}{In fact, one requirement of Weinberg's original proof is that
the momenta are defined in an integer-dimensioned vector space; the extension
to non-integer $d$ has not been established to my knowledge. There are two
responses: either set $d=3$ at this stage to yield a ``physical'' theory,
or note that it is always possible to route a simple loop momentum through
any internal auxiliary line, in which case it may be possible to analytically
continue the integrand to $d=3$ where the theorem holds. We shall not
pursue these rather abstract issues further.}

To the reader familiar with the calculation of the two-loop vacuum
polarization in ${\rm QED}_4$ (eg [21] ch. 8), it is worth making a further
comment. The analogue of the $k^{d-2}\ln\Lambda$ diveregnces are {\sl not\/}
(as one might first think) the $k^2\ln\Lambda$ terms which lead
to a physical charge renormalization -- these are polynomial in $k$
and ``belong'' to the diagram as a whole -- but rather the non-polynomial
$\ln k\ln\Lambda$ divergences, which in general are cancelled by counterterm
subtractions, and in the particular case of QED cancel between the two diagrams
(4a) and (4b). The discontinuous behaviour as $d\to4$ illustrates why this
is a critical dimension for the model.

So, we see that in a sense the main achievement of this paper is simply
the verification of a peculiar case of Weinberg's theorem. There appears
to be no fundamental obstacle to formulating a proof of renormalizability
of the $1/N_f$ expansion for four-fermi theories to all orders. The only
complication arises, as we have seen, because graphs with different
numbers of loops arise at a given order, which means that counterterm
subtractions {\sl of the same order\/} must be applied to yield a finite
result, or in the language of [8], non-trivial cancellations between
divergent graphs at the same order must occur. As we argued in the last
section,
this does not occur in the $1/N_f$
expansion of the nonlinear sigma model.

We deduce that if the expansion is renormalizable, the logarithmic
corrections to fermion and vector propagators and the vertex always
cancel at each order in $1/N_f$.
An important physical consequence, which also follows from Weinberg's
theorem, is that in both
Gross-Neveu and Thirring models, the amplitude for four fermion
scattering receives no $1/N_f$ corrections in the deep Euclidean limit
[8]. In both cases it assumes a universal form $A_d/(N_fk^{d-2})$,
which thus characterizes an interacting ultraviolet fixed point of the
renormalization group. Both models resemble each other at high energies.
At low energies they differ, and now we return our focus to the Thirring
model. In section II we saw that at leading order the coupling $g$ is
completely unconstrained, and that the model can be formulated either as
weakly or strongly coupled. After radiative corrections, this may no
longer be true. Due to the mass renormalization (3.14), the fermion mass
operator acquires an anomalous dimension of $O(1/N_f)$. The result is
that
$${m\over M}\propto\left({\Lambda\over m}\right)^{{C_d\over N_f}
{{2(d-1)^2}\over{d(d-2)}}};\eqno(4.2)$$
that is, for fixed physical mass $m$, the bare mass $M$ must be tuned to
zero as the cutoff is removed. Accordingly the ratio $g^2/M^{2-d}$,
which governs whether the model is weakly or strongly coupled at leading
order, must grow small. However, since the low energy nature of the
model is characterized by the ratio $M_V/m$, it will be necessary to
compute $O(1/N_f)$ corrections to $M_V$ to determine whether the model is
driven to strong or weak coupling at higher orders.
It would also be interesting to test the stability of (4.2) under
corrections of $O(1/N_f^2)$.

Finally we speculate on the relevance of this model to strongly-coupled
QED, both in 3 and 4 dimensions. This paper has been concerned
exclusively with $1/N_f$ perturbation theory. In [19,23],
the leading order vector propagator (2.15) was used in the
Schwinger-Dyson equation to solve for dynamical fermion mass generation
self-consistently. The result, $\propto\exp(-N_f)$, is non-perturbative
in $1/N_f$. It is suggested that the solution may shed light on spontaneous
chiral symmetry breaking in ${\rm QED}_3$, which is suspected to be
critically dependent on $N_f$ [24]. Since ${\rm QED}_3$ is
super-renormalizable, the continuum limit is thought to exist in the
limit of weak coupling; so far we can reach no
conclusion for the Thirring model. It is also worth noting that since
all $O(1/N_f)$ corrections to the vector propagator are UV finite, then
in the asymptotic regime the quenched approximation (which must be made
to solve the Schwinger-Dyson equation) is exact. This is therefore a
model with many similarities to QED in which charge screening is
naturally switched off at short distences --
an effect which must be postulated in studies
of a non-trivial fixed point of ${\rm QED}_4$ due
to fermion mass generation at strong coupling. This
deserves further study. Another possibility is to introduce both
scalar and vector current interactions with independent couplings, to
generate a fermionic analogue of the gauged Nambu -- Jona-Lasinio model,
which would be renormalizable in $d\in(2,4)$. It would then be interesting
to examine the limit $d\to4_-$ and compare the results with other
approaches [25].
\vskip 1 truecm
\noindent{\bf Acknowledgements}

\noindent
This research was supported partly by a CERN Fellowship, and partly by a
PPARC Advanced Fellowship. The two-loop integral was performed with the
help of the symbolic manipulation package FORM. \hfill\break
This paper is dedicated to
the memory of Joan Oswald.
\vskip 1 truecm
\noindent{\bf References}

\noindent
[1] K.G. Wilson, Phys. Rev. {\bf D7}, 2911 (1973).

\noindent
[2] D.J. Gross, in ``Methods in Field Theory'' (Les Houches XXVIII) (R.
Balian and J. Zinn-Justin, Eds.), North Holland, Amsterdam, 1976.

\noindent
[3] K.-I. Shizuya, Phys. Rev. {\bf D21}, 2327 (1980).

\noindent
[4] B. Rosenstein, B.J. Warr and S.H. Park, Phys. Rev. Lett. {\bf62},
1433 (1989).

\noindent
[5] B. Rosenstein, B.J. Warr and S.H. Park, Phys. Rep. {\bf205},
59 (1991).

\noindent
[6] Y. Kikukawa and K. Yamawaki, Phys. Lett. {\bf B234}, 497 (1990).

\noindent
[7] S.J. Hands, A. Koci\'c and J.B. Kogut, Phys. Lett. {\bf B273}, 111
(1991).

\noindent
[8] S.J. Hands, A. Koci\'c and J.B. Kogut, Ann. Phys. {\bf224}, 29
(1993).

\noindent
[9] G. Gat, A. Kovner, B. Rosenstein and B.J. Warr, Phys. Lett. {\bf
B240}, 158 (1990);\hfill\break
H.-J. He, Y.-P. Kuang, Q. Wang and Y.-P. Yi, Phys. Rev. {\bf D45}, 4610
(1992).

\noindent
[10] G. Gat, A. Kovner and B. Rosenstein, Nucl. Phys. {\bf B385}, 76
(1992);\hfill\break
J. Zinn-Justin, Nucl. Phys. {\bf B367}, 105 (1991).

\noindent
[11] J.A. Gracey, Int. J. Mod. Phys. {\bf A6}, 395, 2755(E) (1991);
Phys. Lett. {\bf B297}, 293 (1992).

\noindent
[12] J.A. Gracey, Int. J. Mod. Phys. {\bf A9}, 727 (1994).

\noindent
[13] L. K\"arkk\"ainen, R. Lacaze, P. Lacock and B. Petersson,
Nucl. Phys. {\bf B415}, 781 (1994).

\noindent
[14] K. Symanzik, DESY preprint 77/05 (1977) (unpublished);\hfill\break
I.Ya. Aref'eva, Ann. Phys. {\bf117}, 393 (1979).

\noindent
[15] A.N. Vasil'ev and M.Yu. Nalimov, Theor. Math. Phys. {\bf55}, 163
(1983).

\noindent
[16] G. Parisi, Nucl. Phys. {\bf B100}, 368 (1975).

\noindent
[17] S. Hikami and T. Muta, Prog. Theo. Phys. {\bf57}, 785 (1977).

\noindent
[18] Z. Yang, Texas preprint UTTG-40-90 (1990) (unpublished).

\noindent
[19] M. Gomes, R.S. Mendes, R.F. Ribeiro and A.J. da Silva, Phys. Rev.
{\bf D43}, 3516 (1991).

\noindent
[20] S. Weinberg, Phys. Rev. {\bf118}, 838 (1960).

\noindent
[21] C. Itzykson and J.-B. Zuber, {\sl Quantum Field Theory\/}, Mc-Graw
Hill Book Co., (1980).

\noindent
[22] M. Abramowitz and I.A. Stegun, {\sl Handbook of Mathematical
Functions\/}, Dover Publications Inc., New York (1965).

\noindent
[23] D.K. Hong and S.H. Park, Phys. Rev. {\bf D49}, 5507 (1994).

\noindent
[24] R.D. Pisarski, Phys. Rev. {\bf D29}, 2423 (1984);\hfill\break
T. Appelquist, D. Nash and L.C.R. Wijewardhana, Phys. Rev. Lett.
{\bf60}, 2575 (1985).

\noindent
[25] K.-I. Kondo, H. Mino and K. Yamawaki, Phys. Rev. {\bf D39}, 2430
(1989);\hfill\break
A. Koci\'c, S.J. Hands, J.B. Kogut and E. Dagotto, Nucl. Phys. {\bf B347},
217 (1990);\hfill\break
W.A. Bardeen, S.T. Love and V.A. Miranskii, Phys. Rev. {\bf D42}, 3514
(1990).
\vfill\eject
\input psfig
\vfill

\vbox{
\centerline{
\psfig{figure=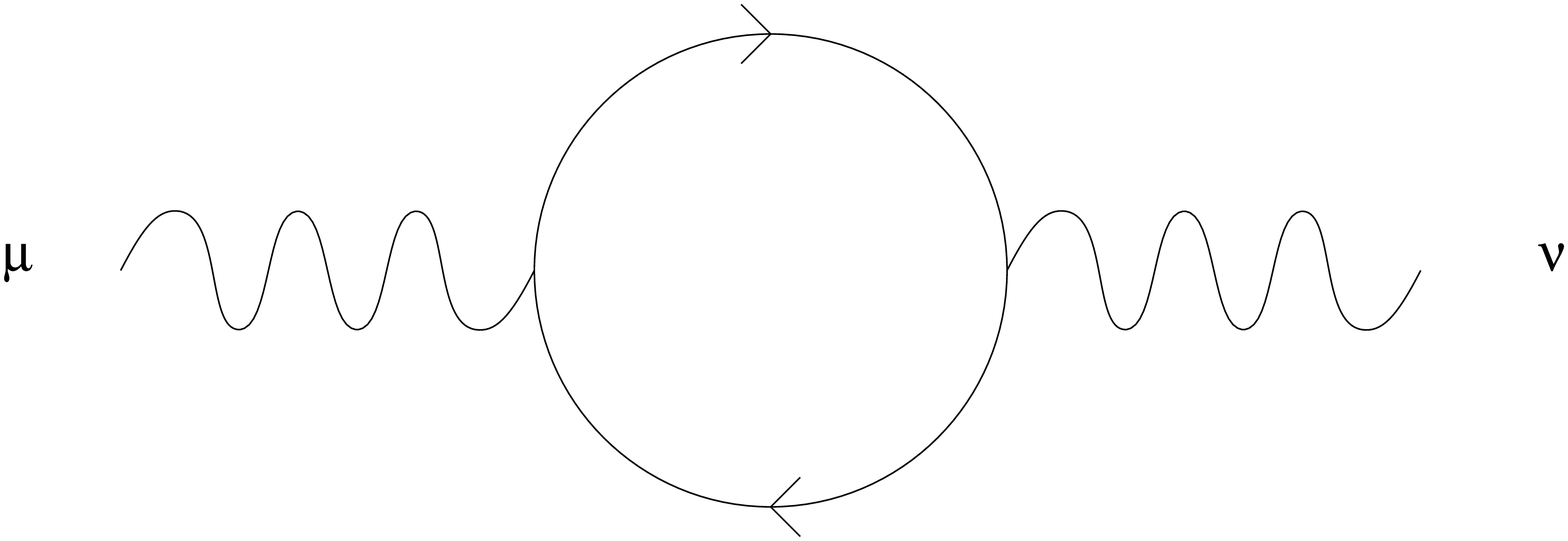,width=5in}}
\bigskip\bigskip
\centerline{\bf Figure 1}
\centerline{Leading order contribution to the vector auxiliary two-point
function}
\centerline{Full lines represent fermions, wavy lines the vector
auxiliary}
}
\vfill
\vbox{
\centerline{
\psfig{figure=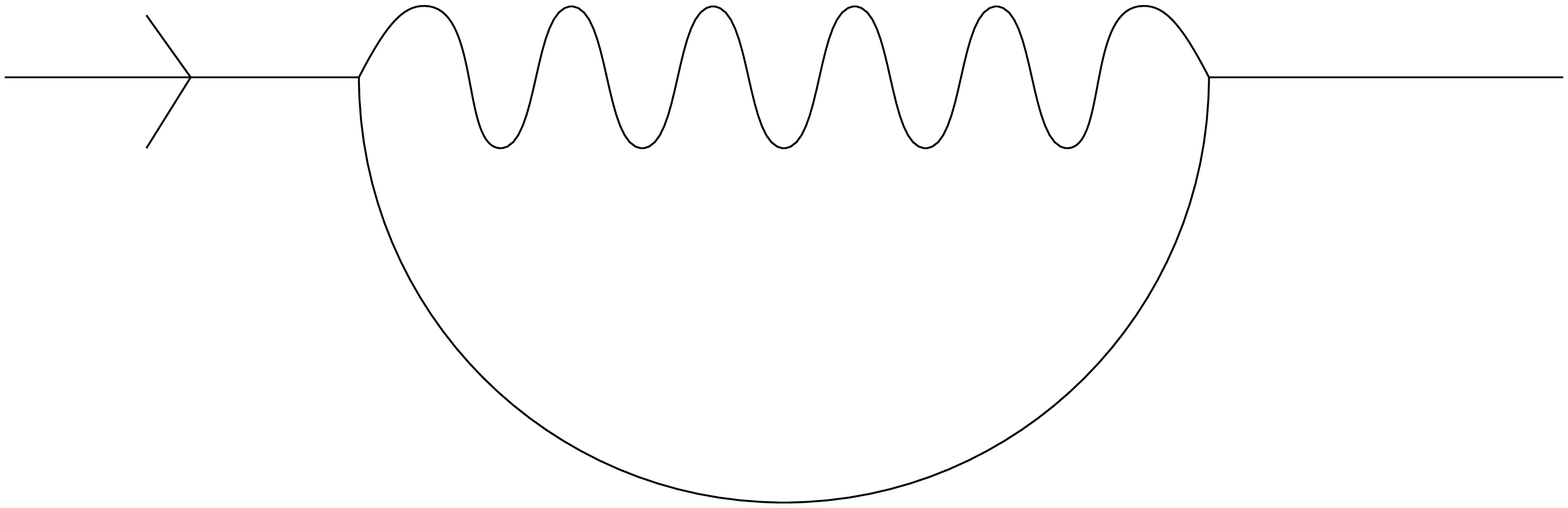,width=5in}}
\bigskip\bigskip
\centerline{\bf Figure 2}
\centerline{$O(1/N_f)$ contribution to the fermion self-energy}
}
\vfill\eject
\vfill
\vbox{
\centerline{
\psfig{figure=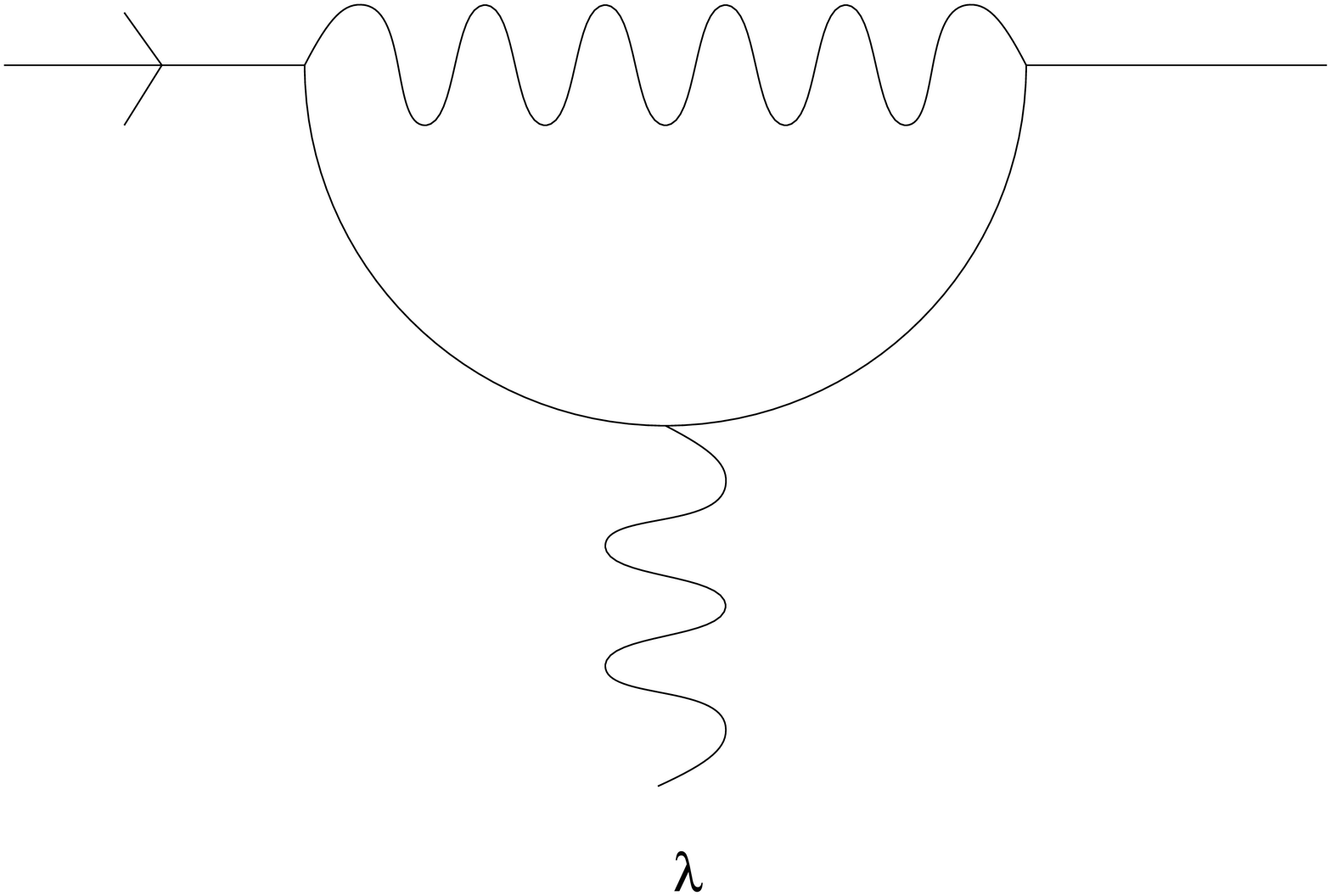,width=5in}}
\bigskip\bigskip
\centerline{\bf Figure 3}
\centerline{$O(1/N_f)$ contribution to the vertex}
}
\vfill\eject
\vfill
\vbox{
\centerline{
\psfig{figure=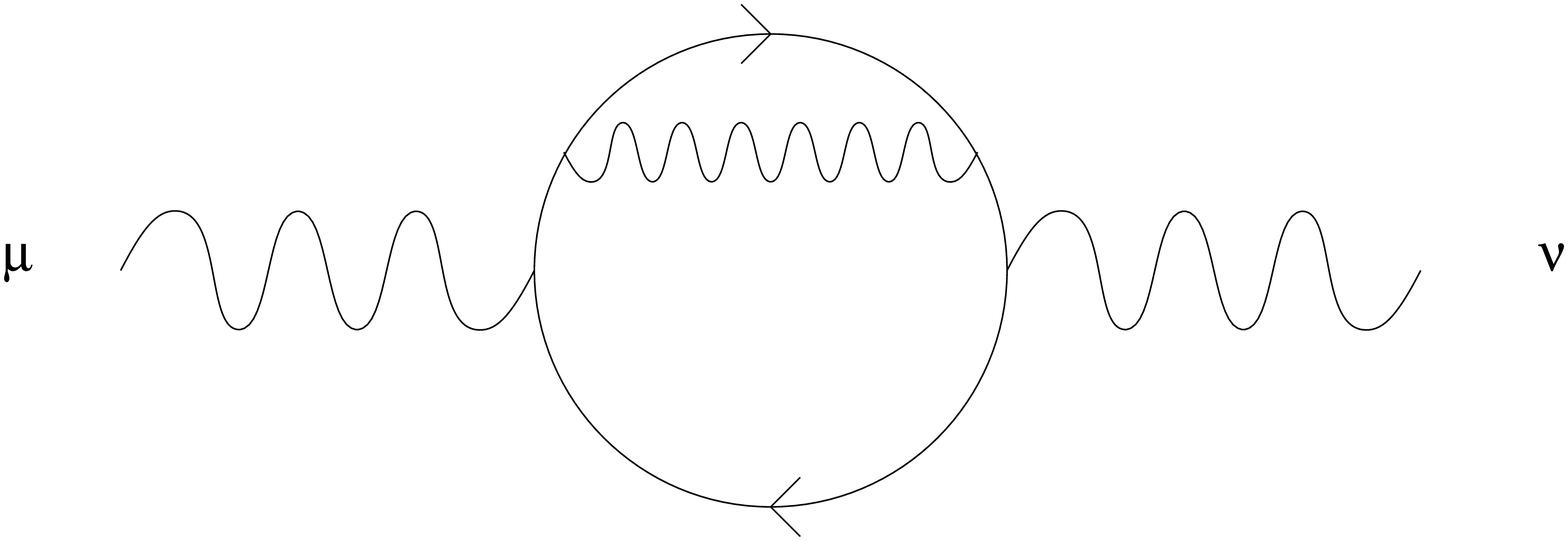,width=5in}}
\bigskip\bigskip
\centerline{\bf Figure 4a}
\centerline{$O(1/N_f)$ contribution to the vector two-point function}
}
\vfill
\vbox{
\centerline{
\psfig{figure=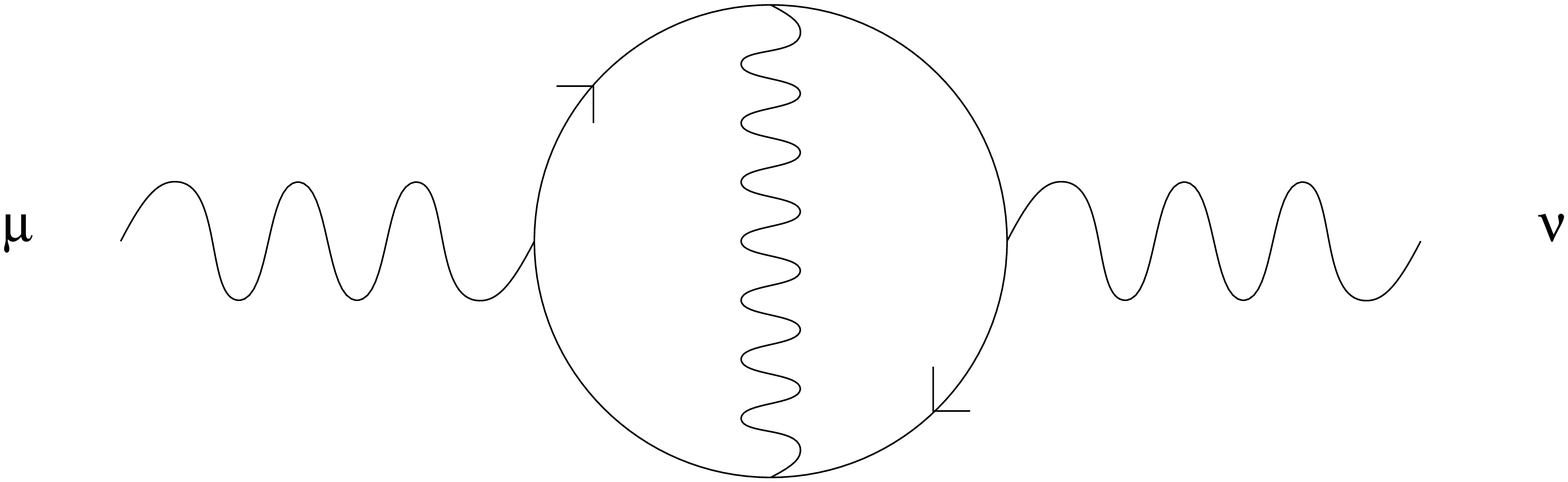,width=5in}}
\bigskip\bigskip
\centerline{\bf Figure 4b}
\centerline{$O(1/N_f)$ contribution to the vector two-point function}
}
\vfill\eject
\vfill
\vbox{
\centerline{
\psfig{figure=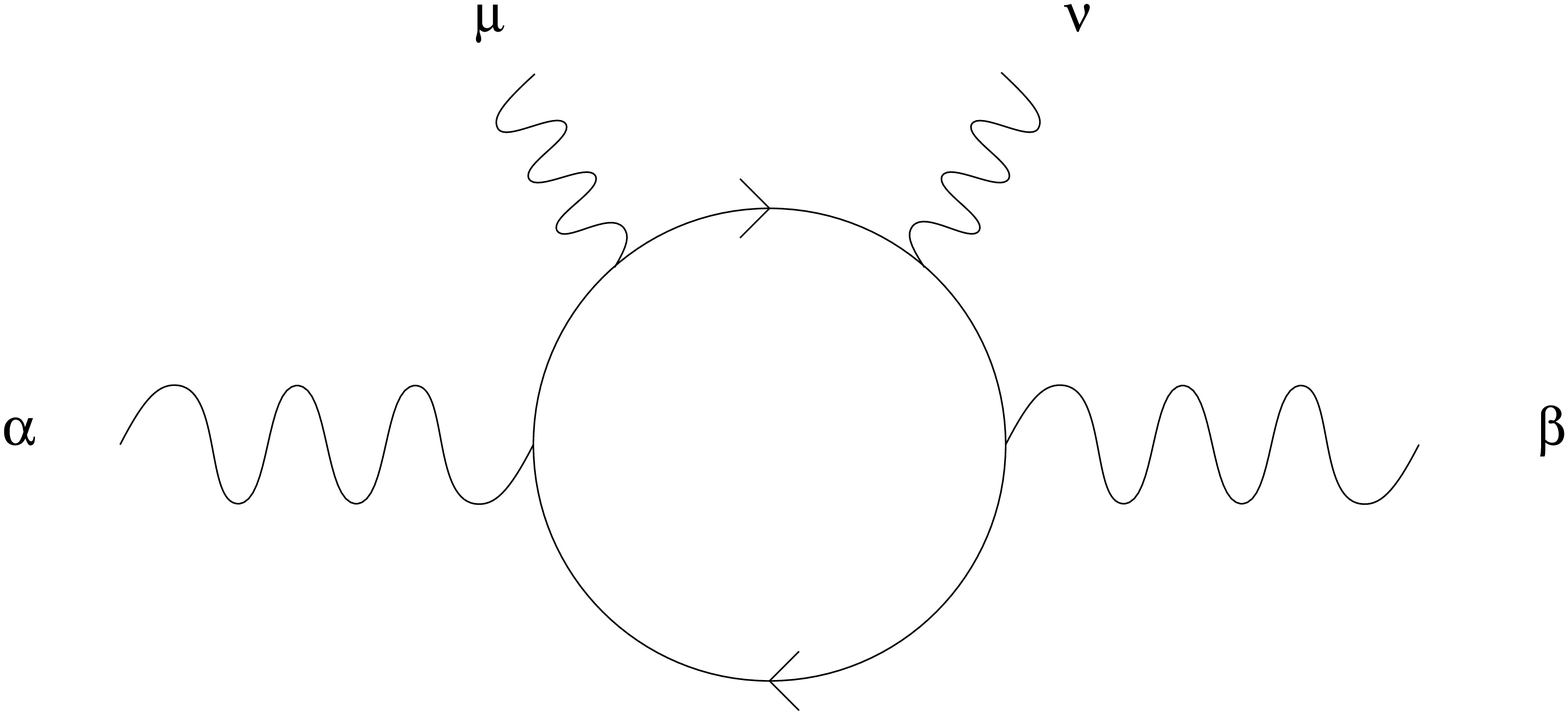,width=5in}}
\bigskip\bigskip
\centerline{\bf Figure 5a}
\centerline{Diagram representing $J_{\mu\nu\alpha\beta}^a(k)$}
}
\vfill
\vbox{
\centerline{
\psfig{figure=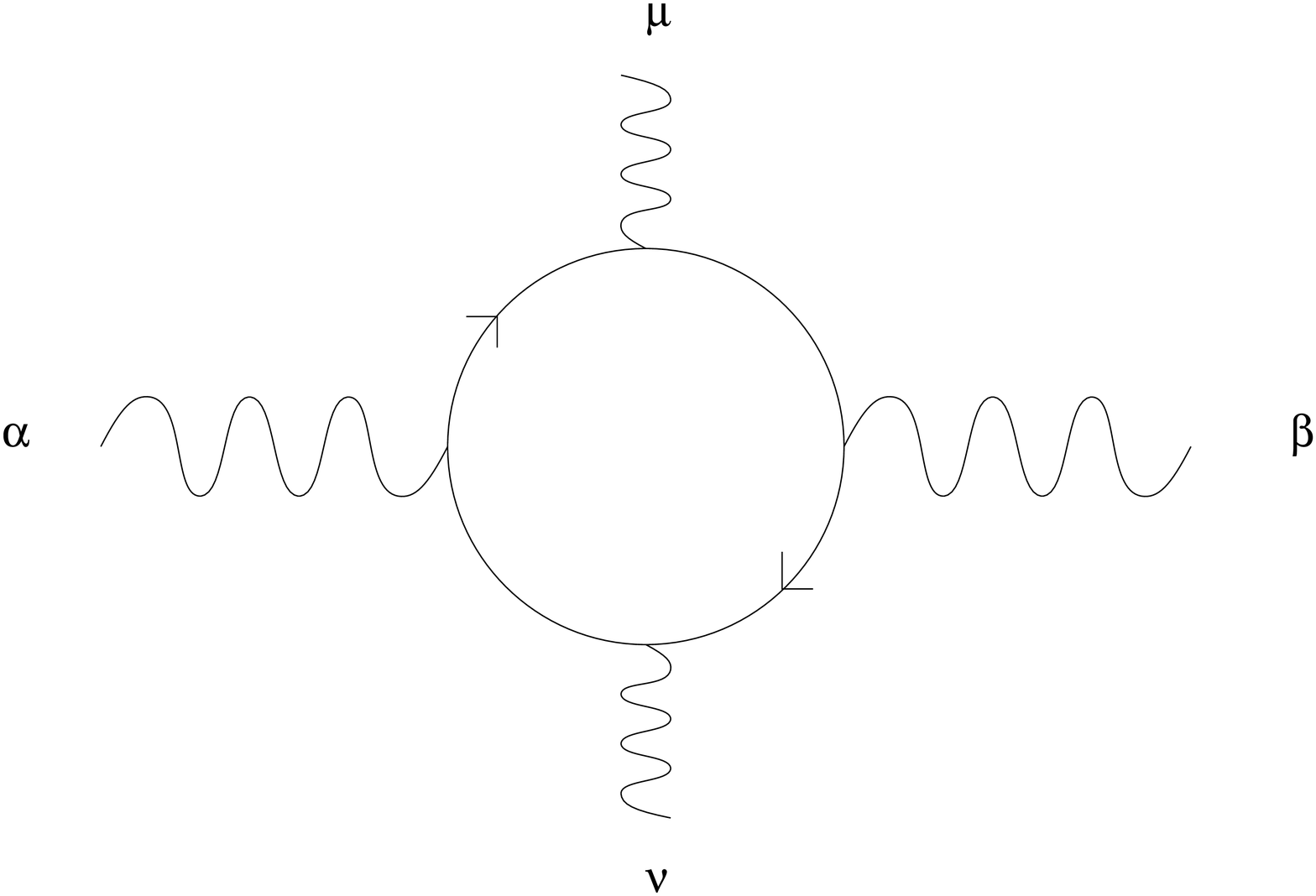,width=5in}}
\bigskip\bigskip
\centerline{\bf Figure 5b}
\centerline{Diagram representing $J_{\mu\nu\alpha\beta}^b(k)$}
}
\vfill\end